\documentclass{CSITproc}
\usepackage{amssymb}
\usepackage{url}

\newtheorem{theorem}{Theorem} %Change {Theorem} to {Lemma}, {Definition}, etc. - for the lemma, definition, etc...
\newtheorem{problem}{Problem}
\newtheorem{lemma}{Lemma}

\begin{document}

% Title -------------------------------------------------------------------------
\title{On interval edge-colorings of bipartite graphs of small order}
% Title -------------------------------------------------------------------------

% Authors -----------------------------------------------------------------------
\numberofauthors{2} %The number of authors

\author{
\alignauthor Hrant Khachatrian \\ %First author
    \affiliation{Yerevan State University\\ Yerevan, Armenia }
    \email{hrant.khachatrian@ysu.am}
\alignauthor Tigran Mamikonyan\\ %Second author
    \affiliation{Yerevan State University\\ Yerevan, Armenia }
    \email{tmamikonyan@mail.ru}\\
%\alignauthor xxx\\%...
%    \affiliation{xxx}
%    \email{xxx}\\
}
% Authors -----------------------------------------------------------------------

\date{28 September - 2 October, Yerevan, Armenia}
\maketitle

% Abstract of the paper ------------------------------------------------------
\abstract{An edge-coloring of a graph $G$ with colors $1,\ldots,t$ is an interval $t$-coloring if all colors are used, and the colors of edges incident to each vertex of $G$ are distinct and form an interval of integers. A graph $G$ is interval colorable if it has an interval $t$-coloring for some positive integer t.
The problem of deciding whether a bipartite graph is interval colorable is NP-complete. The smallest known examples of interval non-colorable bipartite graphs have $19$ vertices. On the other hand it is known that the bipartite graphs on at most $14$ vertices are interval colorable. In this work we observe that several classes of bipartite graphs of small order have an interval coloring. In particular, we show that all bipartite graphs on $15$ vertices are interval colorable.}
% Abstract of the paper ------------------------------------------------------

% Keywords -------------------------------------------------------------------
% Keywords -------------------------------------------------------------------

% Section --------------------------------------------------------------------
\section{Introduction}
In this paper we consider only finite, undirected graphs, without loops and multiple edges. $V(G)$ and $E(G)$ denote the sets of vertices and edges, respectively. The degree of the vertex $v \in V(G)$ is denoted by $d_G(v)$. The concepts and notations not defined here can be found in \cite{West}. 

A proper edge-coloring of a graph $G$ is a coloring of the edges of $G$ such that no two adjacent edges receive the same color. If $\alpha$ is a proper edge-coloring of $G$ and $v \in V(G)$, then by $S(v,\alpha)$ we denote the set of colors of the edges incident to $v$. The largest color of $S(v,\alpha)$ is denoted by $\overline{S}(v,\alpha)$. A proper edge-coloring of a graph $G$ with colors $1,...,t$ is called an interval $t$-coloring if all colors are used, and for any vertex $v \in V(G)$, the set $S(v,\alpha)$ is an interval of integers. A graph $G$ is interval colorable if it has an interval $t$-coloring for some $t \in \mathbb{N}$. The set of all interval colorable graphs is denoted by $\mathfrak{N}$. 

The concept of interval edge-coloring of graphs was introduced by Asratian and Kamalian \cite{AK1987,AK1994}. In \cite{K1989}, Kamalian proved that all complete bipartite graphs and trees are interval colorable. In \cite{P2010,PKT2013}, it was shown that $n$-dimensional cubes have interval $t$-coloring if and only if $ n \leq t \leq \frac{n(n+1)}{2}$. In \cite{Sevastjanov1990}, Sevastjanov proved that it is an NP-complete problem to decide whether a bipartite graph has an interval coloring or not. 

\begin{figure}[t]
\label{fig19}
\begin{center}
\includegraphics[width=8cm]{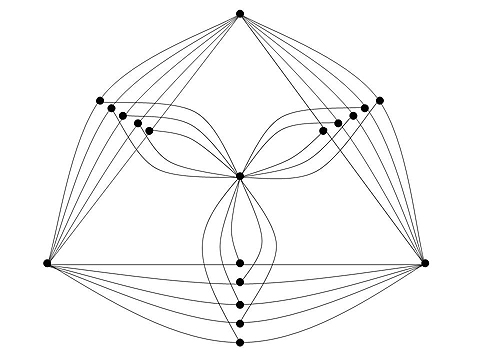}
\caption{Interval non-colorable bipartite graph $F$ on $19$ vertices}
\vspace{1.7cm}
\end{center}
\end{figure}

In \cite{Hansen1992} Hansen proved that every bipartite graph $G$ with $\Delta(G) \leq 3$ is interval colorable. In \cite{JensenToft} Jensen and Toft formulated the following question: is there an interval non-colorable bipartite graph $G$ with $4 \leq \Delta(G) \leq 12$? A partial answer to this question was given by Petrosyan and the first author in \cite{PK2014}. In particular they constructed two interval non-colorable bipartite graphs, $G$ and $H$, where $|V(G)|=20$, $\Delta(G)=11$, and $|V(H)|=19$, $\Delta(H)=12$. Another example of interval non-colorable bipartite graph $F$ on $19$ vertices (Fig. \ref{fig19}) was discovered by Mirumyan in 1989, but was not published, and was independently found by Giaro, Kubale and Malafiejski in \cite{GKM1999}. On the other hand, in \cite{G1999} it was shown that all bipartite graphs on at most $14$ vertices are interval colorable. Based on these results the following problem was posed in \cite{PK2014}:
\begin{problem}
Is there a bipartite graph $G$ with $15 \leq |V(G)| \leq 18$ and $G \notin \mathfrak{N}$? 
\end{problem}

In this work we partially solve this problem by showing that all bipartite graphs on at most $15$ vertices are interval colorable. We also show that some other classes of bipartite graphs of small order are interval colorable.

\section{Related work}
In 1999, Giaro \cite{G1999} used computer search to show that the following result holds. 
\begin{theorem}
\label{Giaro14}
All bipartite graphs of order at most $14$ are interval colorable.
\end{theorem}

In \cite{GKM1999} it was observed that the both parts of an interval non-colorable bipartite graph should be relatively large.
\begin{theorem}
\label{GiaroBipartition}
If $G$ is a bipartite graph with bipartition $(X,Y)$ and $\min\{|X|,|Y|\} \leq 3$, then $G \in \mathfrak{N}$. 
\end{theorem}

Several algorithms for finding a generalized version of interval edge-colorings of graphs are presented and compared in \cite{BHD2009}.

\section {Implementation details}
We use a computer search to look for interval non-colorable graphs of small order. First we generate a set of candidate graphs, then we try to color them using distributed computing system, and finally we double check the obtained colorings for possible errors.

\subsection{Graph generation}
We use the \textbf{nauty} package \cite{McKay} to generate the graphs. In particular, \textbf{genbg} program from the package generates all bipartite graphs according to the given bipartition. Moreover, it allows to specify the minimum and maximum degrees of the graphs, whether the generated graphs should be connected and several other options.

\subsection{Distributed computing}
In order to color the generated graphs we use CrowdProcess, a web-based distributed computing system \cite{CrowdProcess}. CrowdProcess provides an easy to use interface and REST API to submit a program written in JavaScript language together with a list of tasks as a JSON file. It distributes the program and the tasks to multiple computers (between 1000 and 5000 computers during our experiments), gathers the results and passes back through a web interface or an API. Most of the graphs are colored in less than 1 millisecond, so we send up to 200 graphs for each computer.

\subsection{Coloring algorithm}
We represent the bipartite graph $G$ with bipartition $(X,Y)$ and its coloring as a biadjacency matrix $B(G) = (b_{ij})_{n \times m}$, where $|X|=n$ and $|Y|=m$. Here $b_{ij}$ is the color of the edge joining the $i$-th vertex of $X$ and the $j$-th vertex of $Y$, if the edge exists, and is set to $0$ otherwise. We use backtracking to fill in the matrix with colors. At each step we calculate the set of possible colors for the current matrix cell (by taking into account already colored edges). We color the edge by a randomly selected color from the set of possible colors and move to the next cell. If for some edge the set of possible colors is empty, we return to the previous edge and change the color (if there still exist a color in the set of possible colors). The algorithm stops when all the edges are colored, or when all the possibilities are tested and the graph has no interval coloring. In practice the latter never happens because CrowdProcess puts a 5 minute time limit on the computation per computer.

\subsection{Verification}
After downloading the colorings from CrowdProcess we use a C++ program to verify the colorings. We discovered one case when the coloring returned from the CrowdProcess contained an error. We do not know how such an error could occur. We use one more C++ program to gather the graphs which were not colored (due to an error in the coloring or timeout) and send them again. We repeat this process until all the graphs are colored.

\section{Results}
Let $\mathfrak{F}$ be some set of bipartite graphs. Denote by $C(\mathfrak{F})$ the set of all connected bipartite graphs from $\mathfrak{F}$ having minimum degree at least $2$ and having at least $4$ vertices on each of its parts. Also let $M(\mathfrak{F})$ be the set of all graphs obtained by taking any graph $G \in \mathfrak{F}$ and removing any vertex from it. The following lemma holds.

\begin{figure}[t]
\label{fig19-2}
\begin{center}
\includegraphics[width=8cm]{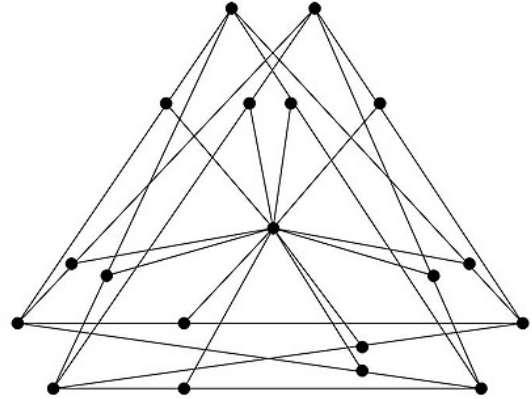}
\caption{Interval non-colorable bipartite graph $H$ on $19$ vertices}
\vspace{1.7cm}
\end{center}
\end{figure}

\begin{lemma}
\label{mainLemma}
If for any set of bipartite graphs $\mathfrak{F}$, all graphs from the sets $M(\mathfrak{F})$ and $C(\mathfrak{F})$ are interval colorable, then all graphs from $\mathfrak{F}$ are also interval colorable.
\end{lemma}
Let $G$ be a bipartite graph from the set $\mathfrak{F}$. If $G \in C(\mathfrak{F})$, then it is interval colorable. Otherwise it is disconnected, has a minimum degree $1$, or has less than $4$ vertices on at least one of its parts. If $G$ is disconnected, then each of its connected components belongs to $M(\mathfrak{F})$, so the union of the colorings of the connected components will be a coloring of $G$. If there exists some pendant edge $uv \in E(G)$, where $d_G(v)=1$, then we take the coloring $\alpha$ of the graph $G-v$ (which belongs to $M(\mathfrak{F})$) and color the edge $uv$ by the color $\overline{S}(u,\alpha) + 1$. Finally, if one of the parts of $G$ has less than $4$ vertices, then $G$ is interval colorable by the Theorem \ref{GiaroBipartition}.
\endproof

\begin{theorem}
\label{th15}
All bipartite graphs on $15$ vertices are interval colorable.
\end{theorem}
Let $\mathfrak{F}$ be the set of all bipartite graphs on $15$ vertices. All graphs from the set $M(\mathfrak{F})$ are interval colorable due to the Theorem \ref{Giaro14}. According to the Lemma \ref{mainLemma} it is sufficient to show that all graphs from the set $C(\mathfrak{F})$ are interval colorable. The number of graphs in the set $C(\mathfrak{F})$ is  288 643 868. We color them all by using a computer algorithm described in the previous section. Some details of the performed computation are presented in Table \ref{table15}.
\endproof

\begin{theorem}
\label{th4x}
All bipartite graphs having $4$ vertices on one part and up to $15$ vertices on the other part are interval colorable except for the graph $G_1$ in Fig. \ref{fig19}.
\end{theorem}
Let $\mathfrak{F}_{i,j}$ be the set of all bipartite graphs with bipartition $(X,Y)$ where $|X|=i$ and $|Y| = j$, $i, j \in \mathbb{N}$. Note that $M(\mathfrak{F}_{i,j})=\mathfrak{F}_{i-1,j} \cup \mathfrak{F}_{i,j-1}$, for any $i,j>1$. We need to prove that all graphs from the sets $\mathfrak{F}_{4,j}$, $12 \leq j \leq 15$, are interval colorable except for the graph $F$ from the Fig. \ref{fig19}. Note that all the graphs from the sets $\mathfrak{F}_{3,j}$ are interval colorable due to the Theorem \ref{GiaroBipartition}. All graphs from the set $\mathfrak{F}_{4,11}$ are also interval colorable due to the Theorem \ref{th15}. We use the computer algorithm described in the previous section to color all graphs from the sets $C(\mathfrak{F}_{4,j})$, $12 \leq j \leq 15$, except for the graph $F$. The details of the computation are presented in Table \ref{table4x}. To complete the proof, we iteratively apply the Lemma \ref{mainLemma} to the sets $\mathfrak{F}_{4,j}$, for $j=12,13,14,15$.
\endproof

\begin{table}[t]
\renewcommand{\arraystretch}{1.2}
\label{table15}
\begin{center}
\begin{tabular}{|c|c|c|}
\hline
No. of vertices & No. of graphs & CPU hours\\
\hline
4 / 11 & 16308 & 3.04 \\
\hline
5 / 10 & 1583646 & 146.35 \\
\hline
6 / 9 & 43739172 & 340.51\\
\hline
7 / 8 & 243304742 & 15537.42\\
\hline
\end{tabular}
\caption{Details of the computation for coloring bipartite graphs of order $15$. CPU hours are reported by CrowdProcess}
\vspace{0.3cm}
\end{center}
\end{table}

\begin{table}[t]
\renewcommand{\arraystretch}{1.2}
\label{table4x}
\begin{center}
\begin{tabular}{|c|c|c|}
\hline
No. of vertices & No. of graphs & CPU hours\\
\hline
4 / 12 & 29515 & 4.96 \\
\hline
4 / 13 & 51616 & 19.19 \\
\hline
4 / 14 & 87609 & 96.95\\
\hline
4 / 15 & 144766 & N/A \\
\hline
\end{tabular}
\caption{Details of the computation for coloring bipartite graphs having $4$ vertices on one part and $j$ vertices on the other part, $12 \leq j \leq 15$. CPU hours are reported by CrowdProcess}
\vspace{1.7cm}
\end{center}
\end{table}

\section{Future work}
Different algorithms can be tried to find interval colorings of the graphs even faster. In fact, the experiments show that most of the graphs have many different interval colorings and possibly some approximation algorithms will be sufficient to find colorings the easily. So far we have tried algorithms based on simulated annealing with no luck.

Currently we are working on coloring the bipartite graphs on $16$ vertices. The number of such graphs after filtering out easily colorable ones (similar to the case of $15$-vertex bipartite graphs) is 12 322 367 816. We have colored more than 98\% of these graphs, but still it cannot be excluded that there exist interval non-colorable graphs on $16$ vertices.

% Sample Figure ------------------------------------------------------------
%\begin{figure}[h]
%\label{sample-figure}
%\begin{center}
%\includegraphics[width=5cm]{sample.eps}
%\caption{Sample figure}
%\end{center}
%\end{figure}

% Sample Table ------------------------------------------------------------
%\begin{table}[h]
%\renewcommand{\arraystretch}{1.2}
%\caption{Sample table}
%\vspace{1mm}
%\label{sample-table}
%\begin{center}
%\begin{tabular}{|c|c|c|}
%\hline
%Title 1 & Title 2  & Title 3\\
%\hline
%item 1,1 & item 1,2  & item 1,3\\
%\hline
%item 2,1 & item 2,2  & item 2,3\\
%\hline
%item 3,1 & item 3,2  & item 3,3\\
%\hline
%\end{tabular}
%\end{center}
%\end{table}

% Sample Theorem ----------------------------------------------------------
%\begin{theorem}
%This is the sample theorem...
%\end{theorem}
%
%\proof The proof of the theorem.
%\endproof

\section{Acknowledgements}
We would like to thank the CrowdProcess team for donating practically unlimited computational time to us and for the excellent technical support. Also we would like to thank UtopianLab coworking space for providing broadband internet access. Finally, we would like to thank P. A. Petrosyan for many valuable comments and ideas. This work was made possible by a research grant from the Armenian National Science and Education Fund (ANSEF) based in New York, USA.

% References -------------------------------------------------------------

% References -------------------------------------------------------------

\end{document}